**Wind lulls and slews; consequences for the stability of future UK electricity systems**

Anthony D Stephens[1] and David R Walwyn[2]

**Abstract**

   As the UK wind fleet increases in size, wind lulls, and wind slews (rapid reductions in wind generation), will increasingly challenge the stability of the UK electricity system. The paper describes how models based on real time records may be used to investigate the most extreme wind lulls and slews likely to be encountered in future, enabling strategies to be devised to mitigate them. Since solar generation is making an increasingly important contribution to the UK electricity system, it was also necessary to include solar slews in the study. Wind lulls are surprisingly frequent, occasionally lasting a week or more, and are always likely to be beyond the capabilities of stored or imported electrical energy to mitigate them. There is likely to be a continuing need for gas powered generation to mitigate wind lulls, and the models provide a means of calculating the consequent annual carbon dioxide emissions. Currently, Combined Cycle Gas Turbines (CCGTs) provide most of the dispatchable generation needed to compensate for any lack of wind or solar generation. However, CCGTs are not sufficiently fast acting to cope with the wind and solar slews anticipated in future. The paper suggests that a range of already proven fast- acting sources of dispatchable generation, including Open Cycle Gas Turbines (OCGTs), Internal Combustion Gas-Fired Reciprocating engines (ICGRs) and stored electrical energy systems, should be capable of coping with the largest wind and solar slews likely to be encountered up to the year 2035.  Examples are given of recent introductions of these fast-acting sources of generation which, it is suggested, will progressively replace CCGTs as the wind and solar fleets increase in size.  Moreover, we see the pattern of recent investments summarised in the paper, as a good indication of likely future investments, with OCGT investments mainly serving the 440 kV grid, and ICGRs and stored electrical energy more local networks.

**Keywords**

Intermittency, wind lulls, wind and solar slews, fast-acting dispatchable generation, energy storage, gas turbine

[1] Correspondence to tonystephensgigg@gmail.com
[2] Graduate School of Technology Management, University of Pretoria, South Africa

1. Introduction

A Royal Academy of Engineering report on wind energy included a discussion about the possible effects on the stability of future United Kingdom (UK) electricity systems with increasing levels of wind capacity (Royal Academy of Engineering, 2014), concluding that the possible deployment of 50 GW of wind capacity by 2030 would represent *"unprecedented levels in any system and raise serious issues of managing the system"*.  The UK government's target for off shore wind capacity alone is now 50 GW in 2030, and a National Grid Leading the Way scenario for 2035, published in FES 2022 (National Grid Electricity System Operator, 2022), suggested a wind capacity of 113.8 GW in 2035 (78.8 GW off shore and 35 GW on shore).  FES 2024 (National Grid Electricity System Operator, 2024), published in July 2024, reveals little change in the National Grid's anticipated wind capacities for 2035 from that in FES 2022.  Clearly, these levels of wind capacity will raise, as the Royal Academy of Engineering suggested, serious management considerations (Royal Academy of Engineering, 2014).

The objective of this study was to develop methods to model wind lulls, and wind and solar slews, likely to be encountered up to the year 2035, and hence to suggest how their destabilising effects might be mitigated.  Based on his study of the Republic of Ireland electricity system of 2006/7, MacKay predicted that the largest future UK wind slews would have a magnitude of 0.37 times the annual average wind generation (MacKay 2009).  It is likely that, for planning purposes, it will be necessary to investigate a large number of different electricity system scenarios.  The finding that the predictions of wind slews in 2030 and 2035 using MacKay's simple algorithm are similar to the predictions of the highly complex models based on real time historic records suggests that it is justifiable to use MacKay's algorithm.

2. Modelling the Intermittency of Wind and Solar Generation

FES 2022 initially included a steady state energy flow diagram for what was described as the Leading the Way scenario in 2035 (National Grid Electricity System Operator, 2022).  Although the 2035 energy flow diagram is no longer available on the National Grid's web site, it was reproduced as Figure 1 in a paper published by the authors in 2024 (Stephens and Walwyn, 2024).  This energy flow diagram, and subsequent private communications with the National Grid in 2023, established that the National Grid anticipated electrical demand increasing to 54 GW in 2035, largely as a result of the need to power electric vehicles and electrically driven heat pumps.  To supply this demand, 5.5 GW of Nuclear Generation would be given the highest priority access to the electricity system, leaving 48.5 GW available to be provided by wind and solar generation. Electrical demand less nuclear generation is an important model variable which we have called the Headroom, or Hdrm, and is calculated as shown in Equation 1.

$$Hdrm = \ electrical\ demand - nuclear\ generation\ \ldots..1$$

Electrical demand currently varies by typically ± 10 GW a day, but it is anticipated that by 2035 it will be possible to arrange for demand to be constant throughout the day, achieved largely by motivating an anticipated 27 million battery electric vehicle owners to charge their vehicles at a time which suits the electricity system, and those with vehicle to grid capability returning some of their stored energy when demand is high (Stephens and Walwyn, 2020b).

The National Grid's 2035 scenario is a steady state model which assumes that the annual average of 54.2 GW of wind generation and 7.1 GW of solar generation will always be available. In reality, wind and solar generation are extremely variable, so the first objective was to create dynamic models which mimicked the variabilities of wind and solar generation and had the same annual average generations as the National Grid's steady state model.

The UK grid is recorded every 5 minutes, the records being available on the internet (Gridwatch, 2021). Downloading the wind records for the years 2013 to 2016 revealed similar wind histograms for each of the four years (Stephens and Walwyn, 2018), implying that any of these four years' records, suitably scaled, could be used to simulate the likely dynamic range of behaviours of wind generation in future years. Recently, solar generation has become an important component of UK's renewable generation and must therefore be included in any model. As the UK solar generation records only became available on the internet in 2017 (Gridwatch, 2021), it was decided to use the records of this year as the basis for modelling future electricity systems.

Given that Gridwatch records the UK electricity system every 5 minutes, there are 104,832 wind and solar records each year. To ease the problem of downloading such a large amount of data, it was decided to download the 2017 records a week at a time, resulting in the creation of 52 weekly models of identical format. A Compound Model was then created by nesting these 52 weekly models inside a spreadsheet which applied input variables simultaneously to all 52 weekly models and collected the calculations of the weekly models to generate annual averages.

The National Grid's proposed steady state average wind and solar generations in the 2035 scenario are respectively 8.96 times and 6.1 times their values in 2017. Figure 1 is a histogram of wind plus solar generation in 2035, generated by applying these multiples to the historical wind and solar generation records for 2017.

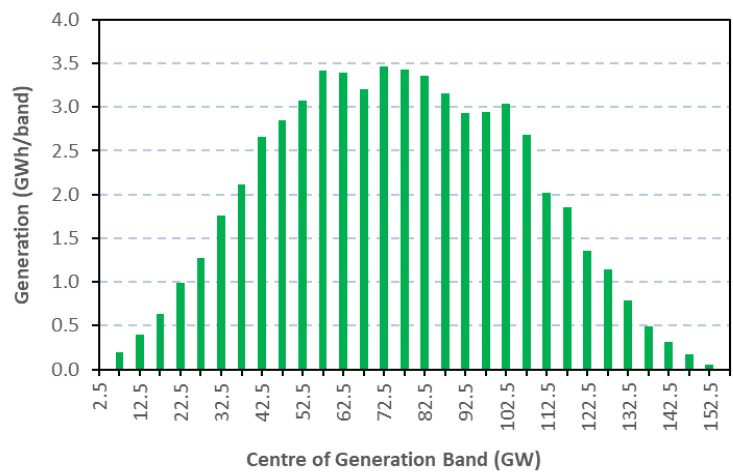

Figure 1. Predicted wind plus solar generation in 2035 in bands of 5 GW

Since a Hdrm of only 48.5 GW will be available in 2035, the question arises of what will happen to wind and solar generation greater than this value, which may be seen in Figure 1 to be up to 100 GW in magnitude. It is sometimes suggested, including by the UK government (Department for Business Energy and Industrial Strategy, 2022), that excess generation could be used to generate hydrogen by electrolysis. However, the Compound Model reveals that excess generation in 2035 will be far too intermittent and variable in magnitude to be put to any beneficial use (Stephens and Walwyn, 2023). Even if that were not the case, it would be far too costly to transmit excess generation, as has recently been confirmed by the Pathway to 2030 plan to spend £10 Bn in transmitting the output of 11 GW of Scottish offshore capacity to England via two 2 GW subsea HVDC cables (Scottish and Southern Electricity News, 2024; New Civil Engineer Newsletter, 2023 ). This means that, when the 11 GW of offshore capacity generates more than 4 GW, the excess will have to be curtailed.

In order to calculate the proportion of curtailment, we use the Compound Model. For every 5-minute interval the 52 weekly dynamic models compare the predicted wind plus solar generation with the available Hdrm, and only generation of Hdrm or less is accepted as dispatched generation. The Compound Model then collects the weekly averages of dispatchable generation needed to satisfy the available Hdrm and the curtailed generation, and hence calculates the annual average values for both outputs.

### 3. Modelling Results

### 3.1 Model Predictions of Future Wind Lulls

Inspection of the 52 weekly dynamic model predictions for 2035 reveal 25 days of wind lulls, including an eight-day wind lull in week 3, three-day wind lulls in weeks 19 and 35, three two-day wind lulls of 15 lulls of a single day. In what follows the implications of the eight-day and three-day wind lulls on system stability will be discussed, together with the German records for the same weeks in 2017, which are available in graphical form on the internet (Fraunhofer ISE).

### 3.1.1    Wind Lull 1 in 2035

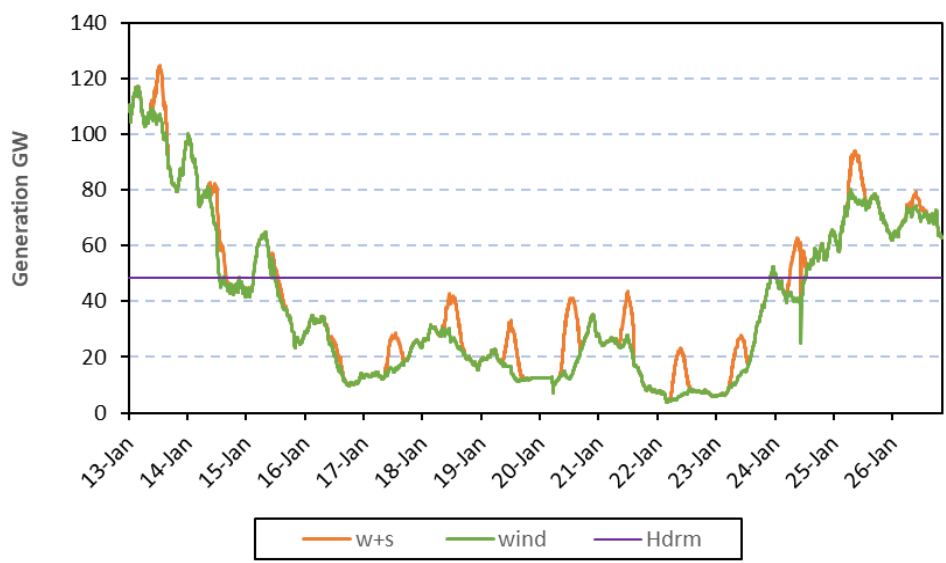

**Figure 2. Prediction of wind and solar generations in 2035 based on eight day wind lull from 16th to 24th January 2017**

Note wind generation is in green, wind plus solar (w+s) in orange and Hdrm is represented by the horizonal line at 48.5 GW

The prediction of minimum wind plus solar generation during the UK wind lull in 2035 is 3.8 GW at 7.20 am on January 22$^{nd}$, leaving 48.5 – 3.8 = 44.7 GW to be provided by dispatchable generation. The overall UK energy deficit over the eight-day wind lull is 5,020 GWh, which may be compared with the total stored electrical energy anticipated in 2035 in FES 2022 of only 150 GWh.

### 3.1.2 Wind Lull 2 in 2035

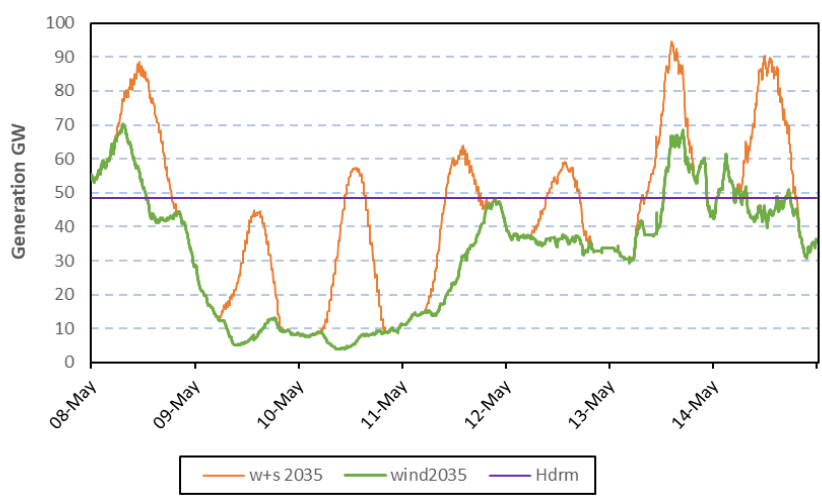

**Figure 3. Prediction of wind and solar generations in 2035 based on three-day wind lull during week 19 of 2017**

The minimum wind plus solar generation in Figure 3 is 7.8 GW, and energy deficit over the three day period is 1,920 GWh.

### 3.1.3 Wind Lull 3 in 2035

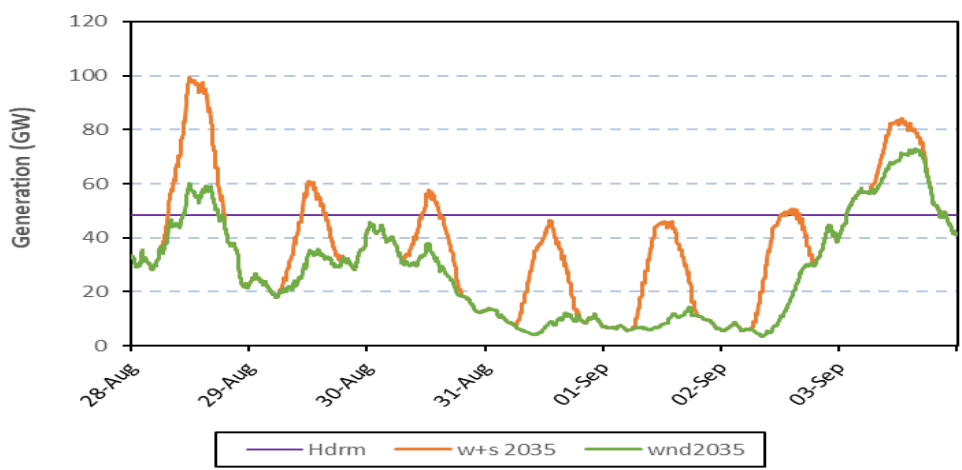

**Figure 4. Predicted wind and solar generations in 2035 based on the 3-day wind lull in week 35 of 2017**

The minimum wind plus solar generation should the wind lull of week 35 2017 reoccur in 2035 would be 5.7 GW at mid-night on September 1, as shown in Figure 4. The energy deficit during the three-day UK wind lull is 1,800 GWh.

The report of the Royal Academy of Engineering (2014) included a comparison of wind generation data in the UK and Germany. Although there was little general correlation between two datasets, levels of wind generation were closely correlated during wind lulls. A comparison of UK wind generations in Figures 2, 3 and 4 with wind records in Germany (Fraunhofer Institute for Solar Energy Systems) reveals coincident wind lulls in the two countries. However, this is not the case for UK wind lulls of less than 3 days duration. An explanation for this finding would appear to be that lengthy wind lulls are a consequence of high-pressure systems being stationary over large geographical areas. Such phenomena are not observed during short wind lulls since these weather patterns tend to be localised. For instance, a weather synoptic for 23rd January 2017 revealed that a lengthy wind lull was experienced across the whole of Europe and the UK, illustrated by a high pressure system with widely spaced isobars sitting over not only the UK but also over most of Europe as far as the Middle East. We note that this synoptic is reproduced in Figure 1 of Stephens and Walwyn (2023).

The importance of the finding that lengthy wind lulls in the UK and Europe coincide is that this undermines the possibility that the 25 GW of interconnectors between the UK and Europe anticipated in 2035 will enable lengthy wind lulls in the UK to be compensated for by imports from Europe. Indeed, a study of the 8-day wind lull in January 2017 revealed an electrical energy deficit in both the UK and Europe (Stephens and Walwyn, 2017). Nor is it possible to assume that wind lulls will be limited to only 8 days. Records reveal a three-week UK wind lull in September 2014 (weeks 36, 37 and 38), as shown in Figure 8 of Stephens and Walwyn (2018), and German energy records confirm that there was an extended wind lull during the same weeks (Fraunhofer Institute for Solar Energy Systems ISE, 2017).

### 3.2 Model Predictions of the Largest Wind and Solar Slews in 2035

Since the dynamic models calculate wind and solar generations, they are also able to calculate their rates of change, $\frac{d(wind)}{dt}$ and $\frac{d(w+s)}{dt}$. Graphs of these rates of change enable the largest slews to be easily identified by searching the weekly dynamic models.

#### 3.2.1 Prediction of Wind Plus Solar Slews in 2035

Week 52 was the windiest in 2017, so we might expect it to produce the most problematic wind and solar slews. Although Figure 5 shows several wind slews of 20 GW/hr or more, all occurred when Hdrm was greater than 48.5 GW. As a result, stability would not have been compromised since excess generation was being curtailed.

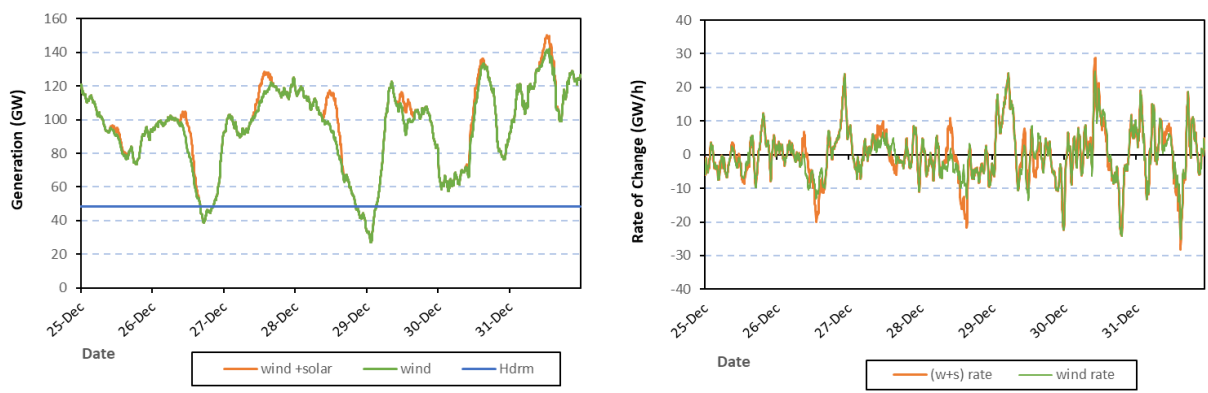

**Figure 5. Prediction of wind and wind plus solar rates of change in 2035 based on records for week 52 2017**

### 3.2.2 Prediction of Largest Wind Plus Solar Slew in 2035

The largest wind plus solar slew in 2035 with the potential to destabilise the electricity system is predicted by the model to occur during week 42, when wind plus solar generation is predicted to fall from 129.7 GW at 12h20 to 8.19 GW at 18h00, almost the entire annual generation range in the histogram of Figure 1. For most of this time wind plus solar generation would be greater than Hdrm, with no threat to stability. However, at 17.05, when wind and solar generation falls to 48.5 GW, the wind slew is 19.19 GW/h and the solar slew 2.74 GW/h, giving a total wind plus solar slew of 21.93 GW/h (see Figure 6). It would be necessary to activate fast acting source(s) of dispatchable generation to maintain stability under these conditions.

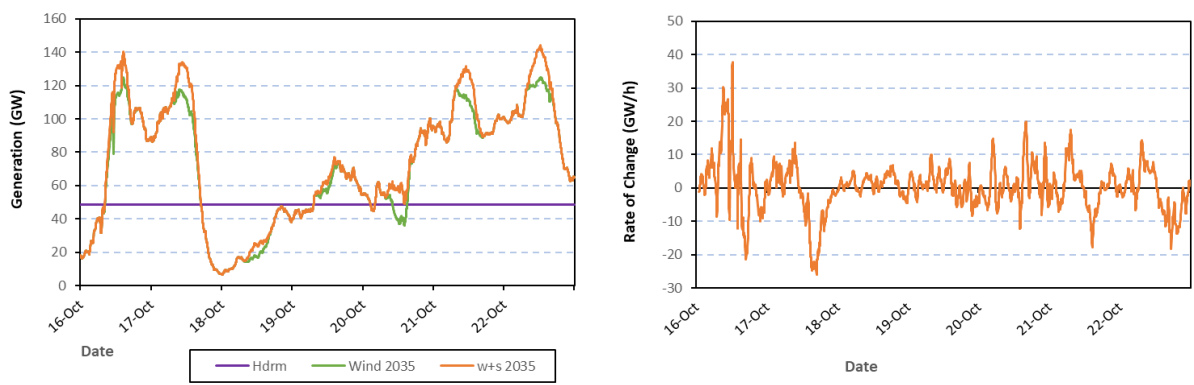

**Figure 6. Prediction of the wind and solar generation slew in 2035 based on week 42 2017 data**

In addition to running the models for the 2035 scenario, they were also run for a hypothetical but realistic 2030 scenario, which assumed Hdrm, wind generation and solar generation values of respectively 35 GW, 26.38 GW, and 3.38 GW. The model predictions of wind slews in 2030 and 2035 appear in Table 1, together with MacKay's prediction that future UK wind slews would be 0.37 times annual average wind generation (MacKay, 2009).

**Table 1. Comparison of predicted wind slews using the dynamic models and MacKay's Rule for the 2035 and 2030 scenarios**

| Scenario | Hdrm | Wind GW | Solar GW | Model prediction of maximum wind slews (GW/h) | MacKay's prediction of maximum wind slews (GW/h) |
|---|---|---|---|---|---|
| 2030 | 35.0 | 26.1 | 3.4 | 10.08 | 9.66 |
| 2035 | 48.5 | 54.2 | 7.1 | 19.19 | 20.00 |

Bearing in mind that wind generation when MacKay studied wind slews in the Republic of Ireland in 2006/7 was only around one hundredth of that predicted for the UK in 2030, it is indeed surprising that MacKay's predictions are so close to the predictions of the Compound Model. A possible explanation for the similarity in the predictions is that wind slews reflect wind duration and speeds, the latter being similar in both countries because they are largely determined by similar meteorological conditions.

There is a large degree of uncertainty about the configuration of future UK electricity systems, with the likelihood that wind slew predictions will have to be frequent assessed as the electricity system becomes progressively better defined. For planning purposes, there is therefore a strong case for adopting the MacKay Rule since it avoids having to resort to complex and time-consuming calculations.

Solar generation has become an important component of UK renewable energy in recent years, making it necessary to consider the destabilising effects of both wind and solar slews. The worst-case scenario from a stability point of view is a maximum wind slew coinciding with a maximum solar slew. Although maximum wind and solar slews never coincided in 2017 this was a matter of happenstance, and the possibility must be entertained in future plans. Luckily, it is a simple matter to calculate what maximum solar slew which must be added to the maximum wind slew. As may be seen in Figure 7, solar generation was at a maximum in week 24, and solar generation and $\frac{d(sol)}{dt}$ (the rate of change in solar generation) for that week are shown in Figure 8, which reveals a maximum solar slew rate of 10 GW/h. This suggests a maximum wind plus solar slew of 30 GW/h in 2035.

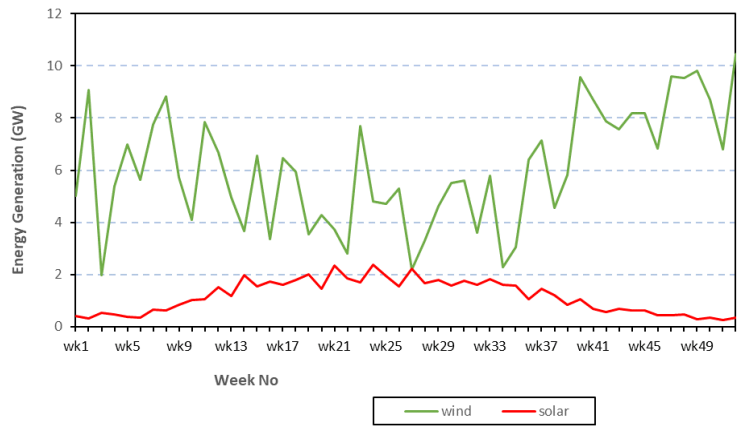

Figure 7. Average weekly wind and solar generations in 2017

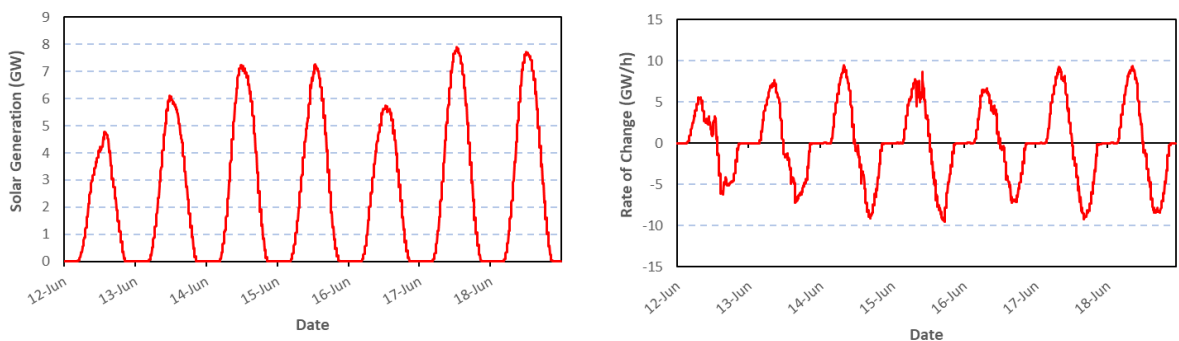

Figure 8. Prediction of solar generation and solar rate of change for week 24 in 2035

### 4. Necessary Measures to Mitigate Wind and Solar Slews

Based on the input values as listed in Table 2, the Compound Model was used to calculate the dispatchable generation needed to satisfy Hdrm and how much had to be curtailed through being in excess of Hdrm. The results are also shown in Table 2 (middle block). The wind slews in the lowest block were then calculated using the MacKay Rule and the likely wind/solar/wind and solar slews by ratioing the 10 GW/h of the 2035 scenario in proportion to the amount of solar generation.

**Table 2. Model predictions for the UK's electricity system outputs and features for different years**

| Year | Units | 2017 | 2023 | 2030 | 2035 |
|---|---|---|---|---|---|
| **Model Input Data** | | | | | |
| Electrical demand | GW | 33.7 | 29.6 | 36.3 | 54 |
| Nuclear generation | GW | 7.48 | 4.37 | 1.3 | 5.5 |
| Hdrm | GW | 26.22 | 25.23 | 35 | 48.5 |
| Wind generation | GW | 6.045 | 9.02 | 26.38 | 54.16 |
| Solar generation | GW | 1.16 | 1.39 | 3.38 | 7.08 |
| | | | | | |
| **Compound Model Predictions** | | | | | |
| Dispatchable generation | GW | 16.0 | 10.2 | 9.99 | 7.44 |
| Curtailed generation | GW | - | - | 4.38 | 20.18 |
| Dispatchable generation as a proportion of Hdrm | GW | 61 | 40 | 28.5 | 15.3 |
| **Slew Calculations** | | | | | |
| Wind slew | GW/h | 2.23 | 3.3 | 9.76 | 20.0 |
| Solar slew | GW/h | c 1.6 | c 1.9 | c 5 | c 10 |
| Wind plus solar slew | GW/h | c 4 | c 5 | c 15 | c 30 |

Generating equipment manufacturers make a wide variety of claims for the performance of their equipment. Table 3 summarises typical claims for close cycle gar turbines (CCGTs), open cycle gas turbines (OCGTs), internal combustion gas reciprocating engines (ICGRs) and stored electrical energy. Not included in the table is pumped hydro power (about 2.7 GW and 30 GWh), which is mainly devoted to meeting rapid increases in demand in the morning and in the evening.

**Table 3. Typical characteristics of main sources of dispatchable generation**

| Parameter | CCGT | OCGT | ICGR | Stored Electrical Energy |
|---|---|---|---|---|
| Typical power | 500 GW | 300 GW | 10-50 MW | 10's of MW |
| Time to full output (minutes) | 20 (from part load) | 2 | 2 | Instantaneous |
| Thermal efficiency at full power | 60 | 35 | 50 | 100 |
| Thermal efficiency at 40% power | 40 | 27 | 40 | 100 |
| Ramp rate (GW/h/GW) | 3 | 30 | 30 | ∞ |

CCGTs have provided most of the dispatchable generation for the UK's system in recent years, although small amounts of OCGT capacity have occasionally been deployed to mitigate rapid changes in demand or wind slews. Such deployments are not easy to identify in the internet records, since they are often ephemeral. A rapid reduction in wind generation on 3rd November 2014 led to the recording of a 0.5 GW of OCGT capacity being a ramped up to full power in 15 minutes before being speedily shut down (see Figure 8 in Stephens and Walwyn (2018) for an illustration of the rapid reduction in wind generation on 3rd November 2014).

Table 3 shows that it will be necessary to substantially increase investment in fast-acting sources of dispatchable generation if wind and solar capacities increase as anticipated in the next decade. Although currently at a relatively low level of capacity compared with the 25 GW of CCGT capacity, an increasing number of investments are now being reported in fast- acting sources of generation, including:

*OCGTs*
- 360 MW of CCGT capacity converted to 245 MW of OCGT capacity at Peterborough in 2018 (Global Energy Monitor, 2023)
- 660 MW of OCGT capacity being commissioned at Kilroot in Northern Ireland in 2024 to replace 700 MW of coal-fired generation decommissioned in 2023 (Irish News, 2024)
- RWE operating a total of 333 MW OCGT capacity in 2024, including 100 MW at Didcot, 53 MW at Hythe, 140 MW at Cowes and 40 MW in Cheshire (RWE, 2024).

*ICGRs*
- 50 MW at Peterborough (Wärtsilä Corporation, 2018)
- 50 MW at Immingham (VPI, 2023)
- 20 MW at Grimsby (VPI, 2024)

*Stored electrical energy systems*
- 49 MW investment by Centrica at Rooscote to increase the stability of the 132 kV South Lakeland District Network (Centrica, 2018)
- 10 MW investment at Kilroot in Northern Ireland in 2024 (Fichtner Consulting Engineers, 2024)

## 5.  Conclusion

The objective of the study described in this paper was to gain an understanding of the frequency and extent of wind lulls and wind and solar slews which have the potential to destabilise future UK electricity systems, and how they might be mitigated. The study made extensive use of a slightly modified Compound Model which had been previously developed by the authors (Stephens and Walwyn, 2020a).  The latter comprises 52 weekly dynamic models of wind and solar generation records for the year 2017, which are then simultaneously scaled by input variables from a covering spreadsheet to simulate wind and solar variabilities in future years.  The covering spreadsheet also collects the calculations of the weekly models to generate annual averages.  The modifications made to the Compound Model for this study were to include in each of the 52 weekly dynamic models calculations of rates of change of wind and solar generation.  Graphing wind and solar generation and their rates of change then allowed wind lulls and wind and solar slews to be identified and studied for future electricity system configurations, with particular reference to electricity system configurations envisaged for 2030 and 2035.

The models revealed frequent wind lulls of one day duration, and less frequent wind lulls of three days and more (the largest being of eight-day duration); it is the lengthy wind lulls which are the biggest threat to the stability of future electricity systems.  Stored electrical energy has been proposed as one means of mitigating wind lulls, but this is not a feasible solution since the stored energy envisaged (150 GWh in 2035) would be exhausted in only a few hours. A recent review of likely battery storage investment worldwide concluded that an anticipated large increase in capacity over the next decade would be driven by the need to power data centres and would be limited in duration to 4-8 hours (Economist, 2024). The typical Li-ion storage cost of $150 per kWh suggested in the article means that an array needed to mitigate the 8-day wind lull of Figure 2, which would only be infrequently used, would cost around $7.5*105 Bn. FES 2022 suggests that 25 GW could be imported from the UK's continental neighbours in 2035. However, this also would not be possible during lengthy wind lulls. They are a consequence of stationary high-pressure systems over not only the UK but also adjacent countries, leading to energy deficits in countries on which the UK has become increasingly reliant for imports. Given our conclusion that these solutions to wind lulls are either impractical or infeasible, it seems unavoidable that the UK will need to retain sufficient gas generating capacity to power its entire electricity system for the duration of lengthy wind lulls (which might last several weeks, as in 2014).

One of the findings of this study was that the prediction of future maximum wind slews up to the year 2035 are similar, whether using our complex model or MacKay's simple prediction that they are likely to be 0.37 times annual average wind generation. This leads us to the conclusion that, for planning purposes, it is more appropriate to use MacKay's simple prediction method, rather than our much more computationally complex method.

A question arises of whether it is possible to deduce the likely main characteristics of future investment in dispatchable generation. In 2017, when dispatchable generation was required to provide 61% of Hdrm, the high thermal efficiencies of CCGTs made them the most appropriate sources of dispatchable generation. However, as may be seen in Tables 2 and 3, thermal efficiency will become less important as wind and solar capacities increase and dispatchable generation as a proportion of Hdrm decreases. The main factors influencing investment in future are likely to be ramp rates, capital costs per GW (rather than per GWh) and local operational considerations.

The list of recent investments listed in Section 4 is likely to be a good indication of the main characteristics of future investments. This suggests that there will be large capacity OCGT investments in generations serving the 440 kV grid, and smaller capacity ICGR and stored electrical energy investments serving more local networks. To mitigate lengthy wind lulls their total capacities will need to be at least equal to the anticipated Hdrm.

### References

Centrica. 2018. *Centrica unwraps Rooscote battery ahead of Christmas Centrica Press release 19 Dec 2018* [Online]. Windsor: Centrica. Available: https://www.centrica.com/media-centre/news/2018/centrica-unwraps-roosecote-battery-ahead-of-christmas/ [Accessed 8 September 2024].

Department for Business Energy and Industrial Strategy. 2022. *British energy security strategy*. London: Gov.UK. Available: https://www.gov.uk/government/publications/british-energy-security-strategy/british-energy-security-strategy [Accessed 29 June 2023].

Economist, T. 2024. *Grid Scale batteries: Charging forward* [Online]. London.

Fichtner Consulting Engineers. 2024. *The 10MW Kilroot Array uses the Advancion® 4 energy storage solution* [Online]. Stockport: Fichtner. Available: https://fichtner.co.uk/projects/kilroot-advancion-energy-storage-array/ [Accessed 8 September 2024].

Fraunhofer Institute for Solar Energy Systems ISE. 2017. *Electricity production in Germany* [Online]. Freiburg: Fraunhofer Available: https://www.energy-charts.de/power.htm [Accessed 4 April 2017].

Global Energy Monitor. 2023. *Peterborough converts 360 MW ccgt converted to 245 MW ocgt* [Online]. London: Global Energy Monitor. Available: https://www.gem.wiki/Peterborough_power_station [Accessed 8 September 2024].

Gridwatch. 2021. *G.B. National Grid Status (data courtesy of Elexon portal and Sheffield University)* [Online]. Templar. Available: http://www.gridwatch.templar.co.uk/ [Accessed 24 January 2022].

Irish News. 2024. *New gas fired turbines go live at Kilroot power station* [Online]. Dublin: The Irish News. Available: https://www.irishnews.com/news/business/new-gas-fired-turbines-go-live-at-kilroot-power-station-I4GKUDSNEVC37A3NWTV3C2XIEI/.

MacKay, D. 2009. *Sustainable Energy - Without the Hot Air*, Cambridge: UIT Cambridge.

National Grid Electricity System Operator. 2022. Future Energy Scenarios (FES 2022). NationalGridESO: London. Available: https://www.nationalgrideso.com/future-energy/future-energy-scenarios. [Accessed 17 May 2023].

National Grid Electricity System Operator. 2024. Future Energy Scenarios (FES 2023). NationalGridESO: London. Available: https://www.nationalgrideso.com/future-energy/future-energy-scenarios-fes. [Accessed 7 September 2024].

New Civil Engineer Newsletter. 2023 *SSEN names contractors to build infrastructure for £10Bn electricity transmission project* [Online]. London: The Civil Engineer. Available: www.newcivilengineer.com/latest/ssen-names-contractors-to-build-infrastructure-for-10bn-electricity-transmission-project-08-08-2023/ [Accessed 8 September 2024].

Royal Academy of Engineering. 2014. Wind Energy Report. Royal Academy of Engineering: London. Available: http://www.raeng.org.uk/publications/reports/wind-energy-implications-of-large-scale-deployment.

RWE. 2024. *OCGT power plants and Gas engine power plants* [Online]. London: RWE in the UK. Available: https://uk.rwe.com/locations/ocgt-power-plants-and-gas-engine-power-plants/ [Accessed 8 September 2024].

Scottish and Southern Electricity News. 2024. *SSEN Transmission's £10bn networks investment to support over 20,000 jobs throughout UK* [Online]. Edinburgh: Scottish and Southern Electricity Network. Available: https://www.ssen-transmission.co.uk/news/news--views/2023/6/ssen-transmissions-10bn-networks-investment-to-support-over-20000-jobs-throughout-uk/ [Accessed 8 September 2024].

Stephens, A. & Walwyn, D. R. 2018. Wind Energy in the United Kingdom: Modelling the Effect of Increases in Installed Capacity on Generation Efficiencies. *Renewable Energy Focus,* 27(Dec 2018)**,** pp 44-58. doi: 10.17632/rs8n356tpk

Stephens, A. D. & Walwyn, D. 2017. The Security of the United Kingdom's Electricity Imports under Conditions of High European Demand. *arXiv,* 1802.07457**,** pp 1-8.

Stephens, A. D. & Walwyn, D. R. 2020a. Development of Mathematical Models to Explore the Potential of Wind Fleets to Decarbonize Electricity Grid Systems. *In:* Maalawi, K. Y. (Ed.) *Modeling, Simulation and Optimization of Wind Farms and Hybrid Systems.* London: IntechOpen, Ch 1, pp 1-19.

Stephens, A. D. & Walwyn, D. R. 2020b. Predicting the Performance of a Future United Kingdom Grid and Wind Fleet When Providing Power to a Fleet of Battery Electric Vehicles. *arXiv preprint arXiv:2101.01065***,** pp 1-18.

Stephens, A. D. & Walwyn, D. R. 2023. Structured Analysis Reveals Fundamental Mathematical Relationships between Wind and Solar Generations and the United Kingdom Electricity System. *arXiv preprint arXiv:2307.11840***,** pp 1-17.

Stephens, A. D. & Walwyn, D. R. 2024. The Development of Investment Planning Models for the United Kingdoms Wind and Solar Fleets. *arXiv preprint arXiv:2403.09496***,** pp 1-16.

VPI. 2023. *VPI expands Immingham energy hub* [Online]. London: VPI. Available: https://vpi.energy/news/archive/vpi-expands-immingham-energy-hub/# [Accessed 8 September 2024].

VPI. 2024. *OCGT power plants and Gas engine power plant* [Online]. London: VPI. Available: https://uk.rwe.com/locations/ocgt-power-plants-and-gas-engine-power-plants/ [Accessed 8 September 2024].

Wärtsilä Corporation. 2018. *Wärtsilä successfully hands over two new plants to support the UK's need for fast-starting flexibility* [Online]. Helsinki: Wartsila. Available: https://www.wartsila.com/media/news/09-10-2018-wartsila-successfully-hands-over-two-new-plants-to-support-the-uk-s-need-for-fast-starting-flexibility-2628166 [Accessed 8 September 2024].